\documentclass[aps,prb,groupedaddress]{revtex4}
\begin{document}
\newcommand{\cm}{cm$^{-1}$}
\newcommand{\nvo}{$\alpha'$-NaV$_{2}$O$_{5}$}
\newcommand{\nxvo}{$\alpha'$-Na$_x$V$_{2}$O$_{5}$}
\newcommand{\avo}{AV$_{2}$O$_{5}$}

\title{Optical properties of $ \alpha '- Na_xV_2O_5$}
\author{M. J. Konstantinovi\'c, Z. V. Popovi\'c, V. V. Moshchalkov,}
\affiliation{Laboratorium voor Vaste-Stoffysica en
Magnetisme, Katholieke Universiteit Leuven, Celestijnenlaan 200D,
B-3001 Leuven, Belgium }
\author{C. Presura,}
\affiliation {Solid State Physics Laboratory, University of
Groningen, Nijenborgh 4, 9747 AG Groningen, The Netherlands}
\author{R. Gaji\'c,}
\affiliation{Institute of Physics, 11080 Belgrade, P.O.Box 68,
Yugoslavia}
\author{M. Isobe and Y. Ueda}
\affiliation{Institute for Solid State Physics, The Univesity of
Tokyo, 5-11-5 Kashiwanoha, Kashiwa, Chiba 277-8581, Japan}

\begin{abstract}
The optical properties of sodium-deficient {\nxvo} ($0.85 \le x
\le 1.00$) single crystals are analyzed in the wide energy range,
from 0.012 to 4.5 eV, using ellipsometry, infrared reflectivity,
and Raman scattering techniques. The material remains insulating
up to the maximal achieved hole concentration of about 15 $\% $.
In sodium deficient samples the optical absorption peak associated
to the fundamental electronic gap develops at $\sim$ 0.44 eV. It
corresponds to the transition between vanadium d$_{xy}$ and the
impurity band, which forms in the middle of the pure {\nvo} gap.
Raman spectra measured with incident photon energy larger then 2
eV show strong resonant behavior, due to the presence of the
hole-doping activated optical transitions, peaked at $ \sim$ 2.8
eV.
\end{abstract}

\maketitle

\section {Introduction}
$\alpha'$- Na$_x$V$_2$O$_5$ belongs to the family of {\avo} oxides
(A=Li, Na, Ca, Mg) which have demonstrated, due to their peculiar
crystal structures \cite{1,1a}, a variety of low-dimensional
phenomena . The nominally pure {\nvo} is a mixed valence compound
($V^{4+}:V^{5+}=1:1$), with the structure consisting of
vanadium-oxygen (VO$_5$) pyramids that are connected via common
edges and corners to form layers in the ({\bf ab}) plane. Its
structure can be described as an array of parallel ladders running
along the {\bf b}-axis, Fig. 1. Each rung is made of a V-O-V bond,
and contains one valence electron donated by the sodium atoms
which are situated between layers. The sodium deficiency does not
alter $\alpha$' crystal structure \cite{1b} up to x=0.70, but
changes the relative abundance of the $V^{4+}$ and $V^{5+}$ ions,
e.g. introduces "extra" holes (more $V^{5+}$ ions) in the ({\bf
ab}) planes.

{\nvo} is one-dimensional antifferomagnetic (AF) insulator at room
temperature \cite{1b}, despite the fact that the vanadium atoms
have uniform valence of $+4.5$, which indicates the quarter-filled
band structure, and suggests a metallic state. However, the
optical measurements \cite {2} revealed the electronic gap of the
order of 1 eV, which is formed between linear combinations of the
$3d_{xy}$ states of the two vanadium ions forming the rungs
\cite{3}. The insulating ground state is argued to originate from
the strong electron-electron interaction \cite {3,3a} e.g. due to
the presence of a large on-site Hubbard repulsion parameter U.

The effects of the hole doping on the optical properties of {\nvo}
have not been studied in details. This interesting topic certainly
deserves attention because of the spectacular effects observed in
doped AF insulators, e.g. high-temperature superconductivity.
Reduction of the Na concentration with respect to the nominally
pure {\nvo} has a strong influence on the room temperature Raman
spectra \cite {4,4a}. The largest effect is the energy shift
(about 35 {\cm}) of the bond bending ($V-O_3-V$) 448 {\cm} phonon
caused by the Na-deficiency. Such a strong renormalization of the
phonon frequency is conjectured to arise from the delocalization
of the electrons from the ladder legs to the bridge oxygens \cite
{4a}.

Here we further examine the electron-phonon coupling effect in
{\nxvo} by measuring Raman spectra using different laser line
energies. We also analyze the influence of the Na-deficiency on
the optical transitions in the mid-IR, and visible energy range.
First, we show that the phonon energy renormalization is not as
large as reported in Refs. \cite{4,4a}, since the structure around
450-500 {\cm} consists in fact of two modes. Their intensity ratio
changes due to the resonant conditions caused by the activation of
the hole-doping associated electronic band, peaked at $ \sim 2.8$
eV. Second, we discuss the electronic structure of {\nxvo} on the
basis of the infrared (IR) reflectivity and ellipsometric
measurements.

\section {Experimental details}
The present work was performed on single crystal plates of
$\alpha'$-Na$_x$V$_2$O$_5$ (0.85$\leq$x$\leq$1.00) with dimensions
typically about $2 \times 4 \times 0.5 $ mm$^3$ in the {\bf a},
{\bf b}, and {\bf c} axes, respectively. The details of the sample
preparation were published elsewhere \cite {1a}. The ellipsometric
measurements on the (001) surfaces of the crystals, both with the
plane of incidence of the light along the {\bf a} and the {\bf b}
axis. An angle of incidence $\theta$ of $66^o$, was used in all
experiments. Procedure for extraction of the dielectric constants
is described in Ref. (\cite{5}). The infrared measurements were
carried out with a BOMEM DA-8 FIR spectrometer. A DTGS
pyroelectric detector was used to cover the wave number region
from 100 to 700 cm$^{-1}$; a liquid nitrogen cooled HgCdTe
detector was used from 500 to 5000 cm$^{-1}$. Spectra were
collected with 2 cm$^{-1}$ resolution, with 1000 interferometer
scans added for each spectrum. Raman spectra were measured in the
backscattering configuration using micro-Raman system with DILOR
triple monochromator including liquid nitrogen cooled
CCD-detector. The Ar and He-Ne lasers were used as excitation
sources.

\section {Raman spectra}
Fig.2 shows the (aa) polarized Raman spectra of {\nxvo}, measured
at room temperature, for several laser line energies. The spectra
are scaled along the vertical-axis for clarity. The spectra of
nominally pure {\nvo}, and the Na-deficient spectra measured with
514.5 nm laser line show the characteristic features which were
already observed, and discussed in previous reports
\cite{10,10a,10b}. Here, we focus on the 350 - 800 {\cm} energy
range, where three $A_g$ phonon modes at 418 {\cm}, 448 {\cm} and
532 {\cm}, and a broad structure centered at 650 {\cm} are found.
Two phonons at 448 {\cm} and 532 {\cm} are strongly interacting
with underlying continuum, which is manifested through their
asymmetric line shapes. In the Na-deficient samples this effect
seems to be strongly enhanced \cite{4a}. The 448 {\cm} mode
corresponds to the $V-O_3-V$ bond bending vibrations, whereas the
532 {\cm} mode corresponds dominantly to the bond stretching
$V-O_2$ vibrations with a small contribution of the $V-O_3-V$ bond
bending vibrations \cite{101}. The origin of the 650 {\cm}
structure is still under debate. Its possible magnetic origin, due
to the AF coupling of the electrons along the ladder direction
({\bf b}-axis), can be ruled out since the structure is observed
only in the (aa) polarized configuration, and because its energy
does not fit the estimates \cite{102} of the exchange coupling
constant. The electronic band structure calculations \cite
{11,11a} show no evidence for the existence of low lying
electronic states even though both Raman and IR spectra confirmed
their activity. The polaronic scenario, as discussed in Ref. [
\cite{4a}], is also questionable since the maximum of the electron
density of states does not fall above the energy of the highest
frequency phonon mode. Nevertheless, Raman spectra indicate the
existence of the strong electron phonon coupling, that should be
related to the charge transfer of the rung electron. The sodium
deficiency introduces the holes in the rungs, so it is natural to
conjecture that such charge dissonance is responsible for the bond
bending phonon renormalization \cite{4a}.

However, the Na-deficiency has very little or no effect on the
Raman spectra measured with the 638.2 nm line, see Fig.2. The
$V-O_3-O$ bond bending phonon at 448 {\cm} does not exhibit large
energy renormalization, and the 650 {\cm} electronic continuum is
clearly visible even in the sample with the $15 \% $ hole
concentration. This finding suggests that it is necessary to
include the resonant effects into consideration. For example, in
Fig.2b we present the (aa) polarized room temperature Raman
spectra of $\alpha'$-Na$_{0.95}$V$_2$O$_5$ single crystal measured
with several laser line energies. This graph shows a typical
evolution of the Raman spectra in {\nxvo} under strong resonant
conditions. Thus, we fit the Raman spectra using Fano model \cite
{12}, which includes the effects of the interaction between
discrete (phonon) and continuum (electron) states. For a
particular energy range, 350 - 800 {\cm}, we took five oscillators
with 17 parameters. It is important to note, as it will be evident
from our fits, that not all the parameters are crucial for the
observed line shape change. The parameters of the first Lorentz
oscillator at 418 {\cm} are kept constant, and used for the
normalization of the spectra, since this mode does not show any
change of intensity, energy, or half width under resonant
conditions. For the phonons at 448 {\cm} and 532 {\cm} we adopt
the Fano line shapes of the form \cite {12}: $I \sim
(q+\epsilon)^2/(1+\epsilon^2)$, with q being the ratio of the
Raman tensors for discrete vibronic and continuum electronic
scattering. The parameter $\epsilon$ is $(\omega - \omega_p)/
\Gamma$, where $ \omega_p$ and $\Gamma$ represent the real and the
imaginary part of the phonon self-energy, when its interaction
with a continuum is taken into account. The continuum is
represented with a Gaussian. The fifth oscillator appears at about
495 {\cm} and its difference from the 448 {\cm} mode is nicely
seen in the Raman spectra of $\alpha'$-Na$_{0.95}$V$_2$O$_5$,
measured with 488 nm laser line, see Fig.2b. It is difficult to
argue whether 495 {\cm} mode interacts with the continuum or not,
so we used Lorentzian for this mode. The results of the fit are
shown as solid lines in Fig.2. The fits are in all cases
excellent. The values of the parameters for
$\alpha'$-Na$_{0.95}$V$_2$O$_5$ are presented in the Table I.

The major change in Raman spectra under resonant conditions is
described by only four parameters; intensity (I), and q, of the
448 {\cm} mode, intensity of the 495 {\cm} mode, and the $\omega$
of the Gaussian. The values of the $\Gamma$, and
$\omega_p=\omega_M-\Gamma / q$ ($\omega_M$ is the position of the
mode maximum) obtained from these fits are, within experimental
error, the same for all spectra. Since the $\Gamma$ is squared
matrix element of the coupling between continuum and discrete
state, this parameter should have no relationship with scattering
wavelength \cite{11}. This is to be expected for the pure resonant
effect, unless there is a change in the strength of the
electron-phonon interaction, caused by the sodium deficiency.
However, these parameters are also found to be the same for all
samples with different Na concentrations, which indicates that the
electron-phonon interaction is not dependent on the hole
concentration. The parameter q of the 448 {\cm} mode has a strong
dependence on scattering frequency, and its variation increases
with increasing sodium deficiency. Similar increase is also found
for the intensity of the 495 {\cm} mode. The square of the
parameter q represents the ratio of the scattering probability for
discrete state to that of the continuum, hence q can exhibit a
dependence on the scattering frequency if the two processes have
different frequency dependencies. This is confirmed by our fits.
The new mode at about 495 {\cm} is probably bond bending vibration
of the rung without electron, since its energy is close to the
corresponding bond bending vibration in V$_2$O$_5$ (482 {\cm},
Ref. \cite {12n}). Under resonance the change of these parameters
(more precisely 1/q, and/or intensity) can be related to the
change of the optical conductivity or dielectric constant \cite
{13}. In the nominally pure {\nvo}, there is no electronic band in
the visible energy region \cite {2,14}, and the Raman spectra show
no dramatic transformation induced by the measurements with
different laser line energies. However, strong resonant scattering
in the Na-deficient samples should originate from the new
electronic transitions at energies between 2 and 3 eV. This band,
induced by hole doping is found at about 2.8 eV using
ellipsometric measurements, as discussed in the next section.

\section {Ellipsometric and infrared measurements}

The optical conductivity of {\nxvo}, obtained from the
ellipsometric measurements is given in Fig.3. For the nominally
pure {\nvo} the bands at the energies 0.9, 1.2, and 3.2 eV for
$\sigma_a$, and at 1.2, 1.6, and 3.7 eV for $\sigma_b$ are found
in accordance with previous reports \cite {14,14a,14b,14c}. The
sodium deficiency causes the activation of the new optical
transitions at about 2.8 eV in $\sigma_a$ and at 3.2 eV in
$\sigma_b$, see Fig.3. As we have already mentioned, the 2.8 eV
peak is responsible for the resonant behavior of the (aa)
polarized Raman spectra. The intensity of the 495 {\cm} mode, see
Inset Fig.3, and the parameter $q^{-1}$ of the 448 {\cm} phonon,
mimic the change in $\sigma_a$, which reinforces our conclusions
about resonant effects in the Raman spectra.

The essential features of the {\nvo} electronic structure are
formed from oxygen 2$p$ and vanadium $3d$ states. The fully
occupied oxygen 2$p$ states are located about 3 eV below the
lowest of the V$3d$ states split by ligand field \cite{15}. The
lowest state of the $V^{4+}$ 3$d$ manifold, 3$d_{xy}$ state, is
occupied by one electron. The V-O-V rung is formed through the
electron hopping via oxygen $O_3$ states. Due to this coupling,
usually denoted as $t_a$, two 3$d_{xy}$ orbital levels form
symmetric and antisymmetric combinations of levels (bonding and
antibonding levels) that are split by 2$t_a \sim 0.7 $ eV \cite
{11a}. The insulating state appears due to the large on-site
Hubbard repulsion, U, which pushes the Fermi level in the middle
of the gap between bonding and antibonding states \cite{3,3a}.
Thus, the main peak in the optical spectrum along the {\bf a}
direction at 0.9 eV seems to originate from the transitions
between bonding (B) and antibonding (AB) states of the same rung
\cite{5}. The higher energy bands, above 3 eV, are similar to
those found in $V_2O_5$ \cite{16}, and correspond to the $O2p -
V3d$ type of transitions.

The sodium deficiency changes the relative abundance of the
$V^{4+}$ and $V^{5+}$ ions. In fact, introduction of holes leeds
to the formation of rungs without electrons, similar to the
$V^{5+}-O_3-V^{5+}$ bonds in V$_2$O$_5$. Therefore, the optical
transitions above 2 eV should also be similar in V$_2$O$_5$ and
{\nxvo}.  However, previous studies \cite {14,14a,14b,14c} did not
clearly show what kind of the final state is involved in the 3.2
eV ($\sigma_a$) transition, and this turns out to be the most
important information in order to understand the electronic
structure of the Na-deficient samples. If the final state is the
bonding V$3d_{xy}$ state, as described in Ref. \cite {14c}, the
{\nvo} should be in the metallic state, which is in strong
disagreement with the observed insulating behavior. In the
insulating state, the optical transitions between $O2p$ and the
bonding V$3d_{xy}$ states should be forbidden. The next
possibility is that the final state of 3.2 eV peak corresponds to
the nonbonding V $3d_{xy}$ state, which forms when there are two
electrons in the rung. According to the modified Heitler-London
model, proposed in order to account for the observed features in
the optical conductivity spectra of {\nvo} \cite{5}, the
nonbonding band lies in the middle of the bonding - antibonding
gap. The sodium deficiency is expected to drive the Fermi level
inside of the bonding bend, which now activates the transitions
between the $O2p$ and bonding V$3d_{xy}$ states. In this case, the
bonding level is partially filled, so this scenario also rely on
the formation of the metallic state in {\nxvo} which is not
observed.

The IR reflectivity measurements show that the material remains
insulating up to the maximal achieved hole concentration of about
$15 \% $. The optical conductivity, obtained from the
Kramers-Kronig (KK) analysis of the IR reflectivity, is presented
in Fig.4. In both $\sigma_a$ and $\sigma_b$ spectra of {\nxvo} we
found no Drude behavior at low frequencies. Instead, the dominant
feature activated by the Na-deficiency is observed in $\sigma_b$
at about 0.44 eV, with a half width at half maximum (HWHM) of
about 0.25 eV. Note that its energy corresponds to the half of the
energy of 0.9 eV peak.

Clearly, all scenarios discussed so far are not yet adequate to
fully describe the optical transitions and the electronic
structure of the {\nxvo}. We argue below that the electronic
structure based on the four electronic bands can account for the
major features observed in the optical conductivity of {\nxvo}.
Simplified schematic representation of the optical transitions,
and the energy levels in {\nvo}, is shown in the Inset of Fig.3.
In the nominally pure {\nvo}, the arrows in the left side of the
Inset of Fig.3 represent the optical transitions which we have
already discussed. The energy levels of the sodium deficient
samples are shown in the right side of the Inset of Fig.3. In
order to account for the appearance of the new modes at 2.8 eV and
0.44 eV in {\nxvo} spectra, we have to assume the existence of the
impurity band (IB), which forms inside the electronic gap of the
pure {\nvo}. This band is further on considered to be empty, so
the optical transitions, denoted as processes "3" in the Fig.3,
may appear in the spectra. We find strong support for our
assumption in the electronic structure, and the optical spectra
\cite{16} of V$_2$O$_5$. In this compound very narrow electronic
band, which is separated from the rest of the conduction band by
an additional gap, exists at the energy of about 2.5 eV. Thus, it
is possible that impurity band in sodium-deficient samples has its
origin from this level. If so, the optical process "3" corresponds
to the transition between oxygen 2p band and IB. Its energy, $E_3
= E_2 - 1/2 E_1 \sim 2.8$ eV, indicates that the IB forms almost
exactly in the middle of the electronic gap of the pure {\nvo}.
This is further supported by the existence of the mid-IR
absorption peak at about $E_{IB} = 1/2 E_1 \sim 0.44$ eV, see
Fig.4, which corresponds to the transitions between V $d_{xy}$ and
IB states. The nature of the IB determines the electronic
properties of {\nxvo}. By increasing sodium deficiency the 0.44 eV
peak grows in intensity and its width increases, which indicates
that the IB gains dispersion. In order to preserve the
preferential insulating ground state (at least up to 15 $\%$
deficiency) this process is accompanied with the increase of the
gap between vanadium $3d$ bands. This is evident from the energy
increase of the peak "1" in $\sigma_a$. Because of that, and since
this band is indeed very narrow (0.25 eV), IB does not overlap
with the occupied d$_{xy}$ band, and the solution is an insulator
since the Fermi level is placed in the middle of the gap between
bonding $d_{xy}$ states and IB. The 0.44 eV peak is found to be
much stronger in $\sigma_b$ then in $\sigma_a$ spectra. Thus, we
tentatively assigned it as the inter-rung electron transition
along the {\bf b} axis.

\section {Conclusion}
In conclusion, we present the analysis of the optical properties
of {\nxvo} using ellipsometric, IR and Raman scattering
techniques. Despite strong hole-doping the material remains
insulating up to the doping concentration of about 15 $\%$. The
insulating state appears as a consequence of very narrow impurity
band formed in the middle of the electronic gap of the pure
{\nvo}. The optical absorption peak develops at $ \sim 0.44 $, eV
and corresponds to the transitions between vanadium V $d_{xy}$ and
impurity band states. The change of the Raman active multiple
phonon structure upon doping, in the energy range between 450 and
550 {\cm}, is the consequence of the resonance, and is caused by
the activation of the hole-doping associated optical transitions
peaked at about $2.8 $ eV .

{\bf Acknowledgments}

M.J.K. and Z.V.P. acknowledge support from the Research Council of
the K.U. Leuven and DWTC.  The work at the K.U. Leuven is
supported by Belgian IUAP and Flemish FWO and GOA Programs.

\newpage
\begin{table}
\caption{Oscillator fit parameters (in {\cm}).}
\begin{tabular}{cccccc}
\hline \hline $\alpha'$-Na$_{0.95}$V$_2$O$_5$ &  & $\lambda$=632.8
nm & $\lambda$=514.5 nm & $\lambda$=488 nm & $\lambda$=457.9 nm
\\ \colrule
1& $\omega$ & 418 & 418  & 418& 418\\
& I & 1500 & 1400 & 1200& 1500 \\
& w & 12 & 12 & 12 & 12 \\
\colrule
2& $\omega_p$ & 445 & 445 & 445& 445\\
& I & 0.5 & 1.8 & 6.5 & 50 \\
& q & 19 & 10 & 4.5 & 0.5 \\
& $\Gamma$ & 20 & 20 & 20 & 20 \\
\colrule
3&$\omega$ & - & 495 & 495 & 495\\
&I & -& 15100 & 27050 & 46100 \\
&w& -& 90 & 80 & 81 \\
\colrule
4& $\omega_p$ & 530 & 530 & 530 & 530\\
& I & 0.7 & 22.7 & 36 & 17 \\
& q & 17 & 3.5 & 2.8 & 3.7 \\
& $\Gamma$ & 10.5 &10.5 & 10.5 & 10.5 \\
\colrule
5& $\omega$ & 664 & 649  & 646& 621\\
& I & 27700 & 38100 & 50000 & 39000\\
& w & 170 & 165 & 182 & 195 \\
\hline
\hline
\end{tabular}
\end{table}

\clearpage

\begin{figure}
\caption { Schematic representation of the $\alpha'$-NaV$_2$O$_5$
crystal structure in the (001) and (010) planes. Effective
representation of parallel ladders coupled in a trellis lattice is
also shown.} \label{fig1}
\end{figure}
\begin{figure}
\caption {Raman spectra of $\alpha'$-Na$_x$V$_2$O$_5$ measured at
T=300 K using various laser line energies for a) x=1, b) x=0.95,
c) x=0.90, and d) x=0.85. The symbols are experimental data, and
the fits are represented with solid lines.} \label{fig2}
\end{figure}
\begin{figure}
\caption {Optical conductivity of $\alpha'$-Na$_x$V$_2$O$_5$
measured at T=300 K for $E \parallel a$ (lower panel) and
$E\parallel b$ (top panel). The x=1 (solid), x=0.98 (dashed),
x=0.95 (dot), x=0.90 (dash-dot), and x=0.85 (solid). Vertical
doted line is placed at energy which denotes the highest laser
energy used in Raman experiment. Low inset: Intensity of 495
cm$^{-1}$ mode as a function of the laser line energy for various
x. Top inset: Schematic representation of the electronic levels in
$\alpha'$-Na$_x$V$_2$O$_5$.}
 \label{fig3}
\end{figure}
\begin{figure}
\caption {Optical conductivity of $\alpha'$-Na$_x$V$_2$O$_5$
obtained from the Kramers-Kronig analysis of the reflectivity
spectra in $E\parallel a$ (top panel) polarizations, and
$E\parallel b$ (lower panel). Inset: Schematic representation of
the V3$d_{x,y}$-IB transitions.} \label{fig4}
\end{figure}

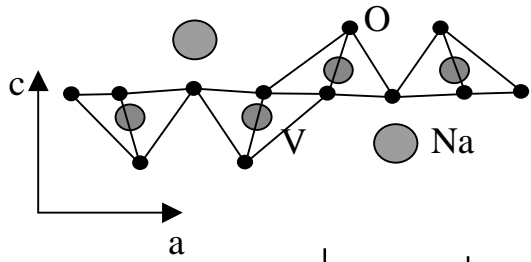
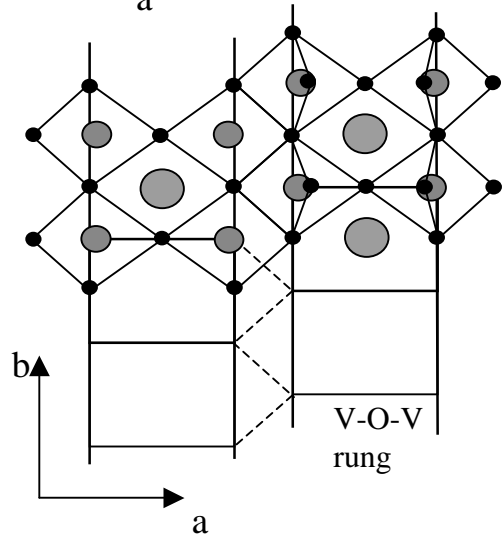
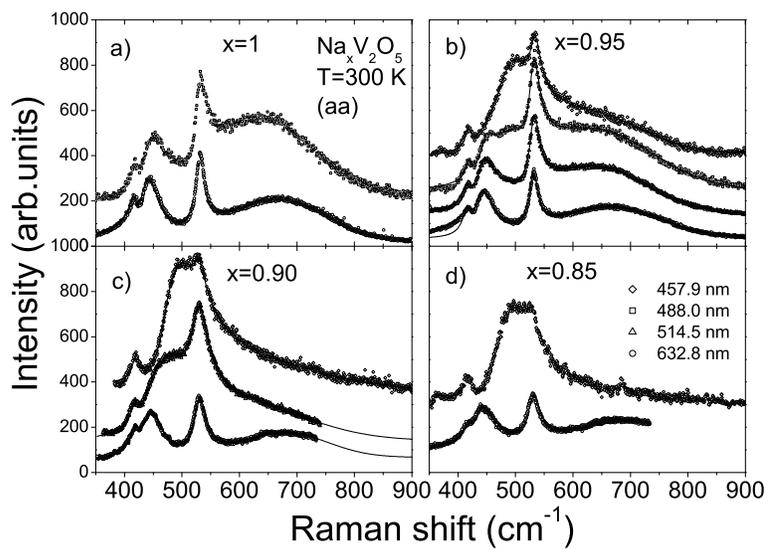
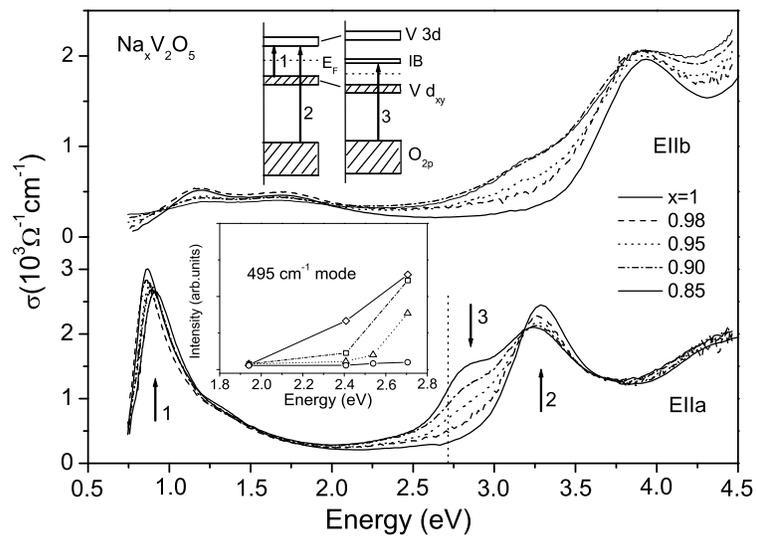
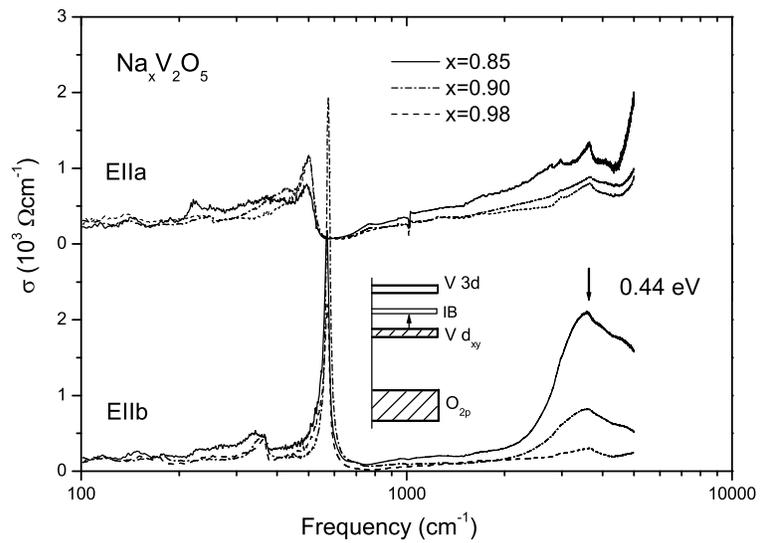
\end{document}